\documentclass[nofootinbib,superscriptaddress,twocolumn]{revtex4-1}
\usepackage{graphicx}
\usepackage[english]{babel}
\usepackage[utf8]{inputenc}
\usepackage[T1]{fontenc}
\usepackage{hyperref}

\usepackage{amssymb}
\usepackage{amsfonts}
\usepackage{amsmath}
\usepackage{color}

\newcommand{\reff}[1]{(\ref{#1})}

\newcommand{\citer}[1]{\cite{#1}}
\newcommand{\citers}[1]{\cite{#1}}

\usepackage{color}

\usepackage{upgreek}

\begin{document}

\title{On the 3D turbulence regime in a Tokamak plasma edge}

\author{G. Montani}
\affiliation{ENEA, Fusion and Nuclear Safety Department, C. R. Frascati,\\
             Via E. Fermi 45, 00044 Frascati (Roma), Italy}       
\affiliation{Physics Department, ``Sapienza'' University of Rome,\\
             P.le Aldo Moro 5, 00185 Roma (Italy)}

\author{N. Carlevaro}
\affiliation{ENEA, Fusion and Nuclear Safety Department, C. R. Frascati,\\
             Via E. Fermi 45, 00044 Frascati (Roma), Italy}


\begin{abstract}
We derive a reduced model for the electrostatic turbulence in a Tokamak edge, when dealing with a resistive plasma and neglecting the spatial gradient of the background density which triggers the linear drift wave response. The obtained dynamics, \emph{de facto} equivalent to a Hasegawa-Wakatani model, is characterized by a constitutive relation between the electric potential Laplacian and the density, which allows to deal with a single 3D equation governing the electric potential fluctuations. We study the evolution of the model, by separating the $n=0$ ($n$ indicating toroidal-like number) mode from all the other ones. Then, we linearize the dynamics of the $n\neq 0$ modes around the steady 2D spectrum, which describes the spectral features of the electrostatic 2D turbulence. We theoretically and numerically demonstrate the existence of decaying branch of the 3D turbulence, having such a 2D steady spectrum as a natural attractor. This result suggests that the basic constituent of the self-sustained non-linear drift response in the plasma edge has to be individualized in the non-linear 2D electrostatic turbulence.
\end{abstract}
\maketitle

\section{Introduction}
In nuclear fusion experiments, like Tokamak devices, a crucial question concerns the energy and particle 
exchange between the magnetically confined 
plasma and the machine walls \cite{wesson}.
In plasma edge, the magnetic 
flux surfaces are open and the plasma has 
a more pronounced collisionality than in the 
machine core. Both from a theoretical \cite{scott07} and 
from an experimental point of view \cite{tcv21,graves05,zweben07}, 
the presence of a plasma turbulent dynamics 
is a well-established feature and it is 
expected to be one of the basic responsible 
for the so-called ``anomalous transport'' 
(i.e. an enhanced transport toward the walls with respect to collisional transport prediction \cite{Braginskii65,mikhailovskii}). 

The turbulence local analysis we develop in the present work mainly concerns the plasma edge dynamics nearby the separatrix. Actually, the study of the scrape-off layer physics would require the introduction of plasma-wall effects represented by specific sheath boundary conditions (see, for example, \citer{stangeby}), not addressed in the present analysis. The origin of the plasma turbulence in the plasma edge has been identified in the non-linear drift wave dynamics, as first clearly shown in \citer{hase-waka83}, and generalized to field curvature effects in \citer{hase-waka87} (more recent review works on these topics can be found in \citers{diamond11,hase-mima18}). A really 3D Hasegawa-Watakani model has been analyzed in \citer{biskamp95} (see also \citers{cheng-okuda77,cheng-okuda78}). This analysis demonstrated that, starting from a random initial configuration, the linear instability drive pumps up the saturation level of the parallel wave numbers which, however, are then suppressed in favour of a nonturbulent poloidal shear flow. This phenomenon is associated the to energy exchange between convective cells and drift waves. In \citer{scott90}, see also \citer{scott02}, it has been clarified how such a turbulence, actually, is not due to the saturation of a linear drift instability, but it is an intrinsically non-linear self-sustained phenomenon. This feature is significantly altered in the external scrape-off layer, where the non-linear instabilities are essentially of ballooning/interchange nature \cite{scott02,garcia2004,garcia2006,peret22}. The linear instability triggers can saturate towards turbulent behaviors leaving a fingerprints on such non linear dynamics \cite{doerk12,navarro15,hatch15}. In \citer{manz18}, it is shown how a notion of universal turbulence can be recovered since such a turbulence can suppress specific features of the unstable linear modes. Moreover, for a discussion of an ingredient which significantly affects the outward transport in the Tokamak edge turbulent regime, see the formation and propagation of so-called ``blobs'' (high density/temperature structures) \citers{boedo14,dippolito2002}. Finally, examples of currently adopted codes to treat plasma edge turbulence in a collisional Braginskii formulation are GBS \cite{ricci-GBS-12}, GRILLIX \cite{stegmeir-GRILLIX-19} and TOKAM3X \cite{tokam3x16}. For completeness, we mention how studies concerning reduced modeling for turbulent phenomena in the core region of the plasma, can be found for example in \citers{sato2022,bernet2015} and references therein.

It is well established (see, for example, the pioneering work \citers{seyler75}, that the basic ingredient of the self-sustained 
turbulence is the non-linear dynamics of 
the electric field. It retains a turbulent character 
also in the 2D case 
(equivalent, in a Tokamak configuration, 
to deal with axial symmetry), when the 
drift coupling term, i.e. the parallel divergence 
of the parallel current density, can vanish. Moreover, 
in \citer{montani-fluids2022}, it was pointed out the existence of an 
exact solution for the vorticity advection 
which corresponds to a flat spectrum in agreement to 
the large wave number components of Kolmogorov-Kraichnan type
\cite{kraichnan80}. In the 2D limit, the plasma 
turbulence is equivalent to that one singled 
out by the Euler equation for an incompressible 
fluid \cite{kraichnan80,seyler75,montani-fluids2022}. Actually, it exists a 
direct mapping between the electric field 
potential and the stream function of the fluid. 
Therefore, it is possible to export to 
the plasma dynamics all the knowledge 
acquired on the fluid turbulence (see, for example, the reviews \citer{boffetta2012} or \citer{zhou21}). In particular, the inviscid dynamics admits (under certain restrictions) 
a statistical equilibrium spectrum determined 
by two constants of motion corresponding 
to the energy and the enstrophy, respectively \cite{kraichnan80}. 
Instead, due to the non conservation of these two constants, the viscous dynamics can be associated to an enstrophy cascade from the smaller to the larger wave number region. Although in \citer{hase-waka83} the 3D nature of the non-linear dynamics is taken into account focusing attention to the most 
linearly unstable mode only, the features 
of the turbulence closely resemble an enstrophy cascade and, hence, 
appear very similar to the 2D case. We remind how, under specific setups, the system can move toward an equilibrium in which most of the energy is singularly condensed in the lowest wave numbers. This phenomenon is called ``condensation'' \cite{kraichnan67,kraichnan75} and it was predicted by using a two-temperature canonical ensemble to describe the spectral distribution for thermal equilibrium. For recent advances on 2D turbulence studies related to a dual cascade mechanism, see \citers{gurcan19,morel21}. While, for critical studies regarding the achievement of the statistical equilibrium, see instead \citer{viviani22,pakter18,dritshel15}.

In this paper, we analyze a reduced description corresponding to the Hasegawa-Wakatani model \cite{hase-waka83} when the background number density gradient and viscosity are neglected and a constitutive relation takes place between the number density fluctuations and the vorticity, i.e. the electric field potential Laplacian. Differently from the original analysis, we consider the full 3D nature of the model and then we linearize the dynamics of the $n\neq 0$ harmonics ($n$ denoting toroidal-like number) around the 2D steady spectrum, corresponding to the analogous fluid dynamics for an incompressible fluid. It is worth remarking that the plasma in the edge region we are considering is not really incompressible but it is the incompressible component itself which drives turbulence. Actually, in what follow we are considering the Boussinesq approximation as applicable everywhere. In particular, we consider the steady solution for the electric potential Fourier dynamics which was derived in \citer{montani-fluids2022} and which corresponds to the statistical equilibrium spectrum when the condensation phenomenon is significantly suppressed. 

We arrive to demonstrate that a decaying branch exists in the $n\neq 0$ harmonics which are exponentially suppressed, so leading to a stability region in the wavenumber space for the 2D turbulence. Then, we validate this theoretical analysis by a numerical integration showing that, when the $n=0$ is initially dominating, all the $n\neq0$ modes decay. This result suggests 
that the self-sustained turbulence associated to the non-linear drift response is strictly correlated to the electrostatic turbulence, which does not 
need the linear triggering of the background number density gradient. 

Moreover, we briefly discuss the spatial 
profile of the number density distribution. We show that, in the Fourier space, the model predicts a constant steady spectrum. As a consequence, when 
we plot the spatial distribution of the 
electric field Laplacian, i.e. of 
the number density apart a constant factor, we see that it is characterized by a large number of small scale 
bumps and depressions. 

Finally, we show how the structure of the dynamics is altered if we restrict our attention to the 2D case introducing the X-point geometry (so that parallel derivatives are now present). In particular, we obtain for such a restated equation an exact
unstable solution, which can be thought as a marker of the peculiar nature
of the turbulence near enough to the X point.

The paper is organized as follows. In Sec.\ref{sec2}, the reduced model for the turbulence is described. The scheme consists of a closed single equation for the electric potential describing the 3D turbulent dynamics of the fluctuations in the absence of ion viscosity and for negligible background density gradient. In Sec.\ref{sec3vs2}, the Fourier expansion the fluctuating potential is addressed and a linearization around the $n=0$ mode is then implemented. The truncated Fourier scheme admits, assuming a constant vorticity steady spectrum for the $n=0$ component, a time decaying turbulence branch for the other $n\neq0$ modes: this indicates how the 2D steady spectrum corresponds to an attractor for the 3D dynamics. In Sec.\ref{secnum}, the non-linearized 3D reduced turbulence equations are numerically integrated in the truncated Fourier space. Assuming a dominant $n=0$ initialization, the decaying behavior of the $n\neq0$ components is clearly outlined. Moreover, it is shown how the constant vorticity spectrum for the $n=0$ component naturally emerges for the steady configuration, indicating the general character of such a solution. Density fluctuations are also plotted outlining small scale structures. In Sec.\ref{secxp}, the effects of the X-point geometry is considered in the limited 2D problem. The presence of a local unstable solution is finally discussed. Concluding remarks follow.

\section{Turbulence reduced model}\label{sec2}
Addressing the scenario of the plasma dynamics 
in a Tokamak edge, we consider a two fluid 
model (ions and electrons) in the quasi-neutrality 
condition and retaining only the parallel electron resistivity 
as non-ideal effect. We adopt Cartesian coordinates 
$(x,y,z)$ and the constant and uniform magnetic field (having intensity $B_0$) is taken along the $z$ axis. Both the species possess the $\textbf{E}\times \textbf{B}$ velocity as the dominant contribution to their motion, while 
the presence of a non-zero current density is ensured 
by the ion drift polarization velocity only. The construction 
of the basic field equations relies on the validity 
of the so-called ``drift ordering'' approximation 
(for more details on the model morphology and its 
relevance for a plasma near the X-point of a Tokamak, 
see \citer{montani-fluids2022}). 
We study the case of a homogeneous and isotropic plasma, 
which background features are characterized by 
a constant (electron) number density $\bar{n}_0$ and 
a constant temperature $T_0$, equal for ions and 
electrons. Thus, all the dynamical quantities 
are here perturbations, denoted by no suffix. 

Considering an Hydrogen-like plasma, the equation 
governing the dynamics of the number density $\bar{n}$ (the same for ions 
and electrons) takes the following form:
\begin{equation}
	\frac{d\bar{n}}{dt} = \frac{1}{e}\partial_zj_z
	\, , 
	\label{ggm1}
\end{equation}
where $e$ is the electron charge, $j_z$ denotes the 
parallel current density and the advective derivative 
is constructed with the $\textbf{E}\times \textbf{B}$ velocity. Expressing the perpendicular current density via the 
ion drift polarization velocity, the charge 
conservation equation (i.e. the divergenceless nature 
of the current density) provides the dynamics 
of the electric field potential $\phi$ as:
\begin{align}
	\frac{d \Delta \phi}{dt} = 4\pi \frac{v_A^2}{c^2}
	\partial_zj_z \, , 
	\label{ggm2}
\end{align}
$v_A$ denoting the Alfv\'en velocity corresponding to 
the magnetic field intensity $B_0$ and $\Delta \equiv \partial_x^2 + \partial_y^2$ is the perpendicular Laplacian. 
Comparing Eq.(\ref{ggm1}) and Eq.(\ref{ggm2}), we easily get the following ``constitutive'' relation:
\begin{align}
	\Delta \phi = 4\pi e \frac{v_A^2}{c^2}\bar{n}
	\, .
	\label{ggm3}
\end{align}
Finally, the parallel electron generalized Ohm law provides the expression 
for the parallel current density in the form:
\begin{align}
	j_z = \sigma \Big(\frac{K_BT_0}{\bar{n}_0e}
	\partial_z \bar{n} - \partial_z\phi\Big)
	\, , 
	\label{ggm4}
\end{align}
here $\sigma$ denotes the parallel 
conductivity coefficient expressed by $\sigma=1.96\, \bar{n}_0 e^2/m_e \nu_{ei}$, where $\nu_{ei}$ is the electron-ion collision frequency ($m_e$ being the electron mass), while $K_B$ is 
the Boltzmann constant and we made use of the perfect gas law. Combining together Eq.(\ref{ggm2}) with Eqs.(\ref{ggm3}) and (\ref{ggm4}), we get the following closed partial differential equation for the electric potential $\phi$:
\begin{align}
&\partial_t\Delta \phi + \frac{c}{B_0}\left( \partial _x\phi \partial_y\Delta \phi - \partial_y\phi\partial_x\Delta \phi \right) = \nonumber\\
&\qquad\qquad\qquad=\frac{\sigma K_BT_0}{\bar{n}_0e^2}\partial^2_z\Delta \phi - 4\pi \frac{v_A^2}{c^2}\sigma \partial^2_z\phi
	\, ,
	\label{ggm5}
\end{align}
where we explicitly expressed the advective derivative.

For the analyzed case, the ion gyrofrequency writes $\Omega_i=eB_0/c m_i$ (with $m_i$ the ion mass), while the ion Larmor radius can be expressed as $\rho_i^2=K_B T_0/m_i \Omega_i^2$ (we note that, since we are assuming equal ion and electron temperatures, $\rho_i$ coincides with the often used quantity $\rho_s$, which is the ion Larmor radius at the electron temperature). Introducing the dimensionless coordinates $\tau\equiv \Omega_i t$, $u\equiv(2\pi/L) x$, $v\equiv(2\pi/L) y$, $w\equiv (2\pi/R) z$ ($L$ and $R$ being two spatial scales, with $R\gg L$) and the dimensionless variable $\Phi\equiv e\phi /K_BT_0$, we can rewrite Eq.(\ref{ggm5}) in the following dimensionless form:
\begin{align}
\partial_{\tau}D\Phi + \bar{\rho}_i^2(\partial_u\Phi\partial_vD\Phi &- \partial_v\Phi  \partial_u  D\Phi) =\nonumber\\
&=(1/\bar{\nu}_{ei})\partial^2_wD\Phi - \gamma\partial^2_w\Phi\,, \label{ggm6}
\end{align}
where
\begin{align}
\bar{\rho}_i=\rho_i\Big(\frac{2\pi}{L}\Big)\,,\quad
\bar{\nu}_{ei}= \nu_{ei}\,\frac{m_e\Omega_i}{K_BT_0}\Big(\frac{R}{2\pi}\Big)^2\,,\;\nonumber\\
\gamma= \frac{\delta_i}{\eta_B / v_A}\,\Big(\frac{L}{R}\Big)^2\,,\qquad\qquad\qquad\quad\quad\;\nonumber 
\end{align}
and $D\equiv \partial^2_u+\partial^2_v$. In the expression of the parameter $\gamma$, we used the definition of ion skin depth $\delta_i=c/\omega_{pi}$ (where $\omega_{pi}^2=4\pi\bar{n}_0e^2/m_i$) and of the magnetic diffusivity $\eta_B=c^2/4\pi\sigma$. 

The present model, summarized by Eq.(\ref{ggm6}) 
can be thought as a simplification of the 
well-known Hasegawa-Wakatani picture of 
the plasma turbulence \cite{hase-waka83}, in which 
the ion viscosity has been removed and the 
background number density gradient (responsible 
for the linear drift instability) has been 
considered as negligible with respect to 
the perturbation gradients. 
Nonetheless, the proposed equation describes turbulence properties of the plasma edge and it contains, 
due to its 3D nature, the non-linear drift response \cite{scott90}. 
In \citer{montani-fluids2022}, it has been emphasized how, 
neglecting the $w$ dependence of the problem, 
the present model is isomorphic to the 
Euler equation for a 2D incompressible fluid \cite{kraichnan80}. The mapping between the two schemes is immediate if the electric potential 
$\Phi$ is identified with the (neutral) 
fluid stream function. Also this 2D reduction is a turbulent system, although the non-linear drift response is absent. 

It is important to stress that
our model refers to Tokamak plasma edge in view
of the basic assumptions on which it is
derived (see \citer{montani-fluids2022}, for further details) and they are essentially the same at the ground of many current edge codes (e.g. \citer{tokam3x16}). The most relevant simplifications concern the choice of the parallel
resistivity (crucial for the drift response) as
the only non-ideal effect and having
neglected the magnetic field curvature (so that we adopt local Cartesian
coordinates). In the following theoretical and
numerical analyses of Sec.\ref{sec3vs2} and \ref{secnum}, the parallel response is along the
$z$-direction and the orthogonal
ones lay in the $(x,y)$-plane
(for a discussion of a more realistic
X-point geometry, see Sec.\ref{secxp}). In the analogy with a Tokamak geometry, the \emph{z}-direction would correspond to the toroidal one while the $(x,y)$ layers to the poloidal slices. Since we are looking at a small region of the plasma near the X-point, we
adopt, in what follows, periodic boundary conditions. This assumption
allows to numerically integrate the
dynamics via a truncated  Fourier approach. However, the theoretical
investigation of the 2D turbulence stability requires only the validity of
a Fourier integral expansion of the
electric field in the orthogonal plane and therefore has a greater degree
of generality than the periodic case.

\section{2D turbulence as attractor for 3D dynamics}\label{sec3vs2}
The $w$ coordinate must reflect the toroidal 
topology of a Tokamak machine, thus we can naturally 
consider $\Phi$ as a periodic function 
of such a variable and, hence, consider 
a Fourier expansion of the form: 
\begin{align}
	\Phi (u,v,w) = \sum_{n=-\infty}^{+\infty}
	\Phi^{(n)}(u,v)e^{inw}
	\, , 
	\label{ccm8}
\end{align}
where $0\le w< 2\pi$ and $n$ is a positive or negative integer. 
Substituting this expansion into Eq.(\ref{ggm6}) 
and separating the equation for $n=0$ from 
those ones for a generic value of $n\neq 0$, 
we get the system
\begin{align}
&\partial_{\tau}D\Phi^{(0)} + 
\bar{\rho}_i^2\big( \partial_u\Phi^{(0)}\partial_vD\Phi^{(0)} 
- \partial_v\Phi^{(0)}\partial_uD\Phi^{(0)}\big) + \nonumber \\
&+ \bar{\rho}_i^2 \sum_{n'\neq 0}\big( 
\partial_u\Phi^{(-n')}\partial_vD\Phi^{(n')} -\partial_v\Phi ^{(-n')}\partial_uD\Phi^{(n')}\big)=0\, , \label{ccm10}
\end{align}
\begin{align}
&\partial_{\tau}D\Phi^{(n)} - n^2\big(\gamma\Phi^{(n)} - (1/\bar{\nu}_{ei}) D\Phi^{(n)}\big)+\nonumber\\
&\qquad+ \bar{\rho}_i^2 \! \sum_{n'=-\infty}^{\infty} 
\big( \partial_u\Phi^{(n-n')}\partial_vD\Phi^{(n')} -\nonumber\\
&\qquad\qquad\qquad\qquad- \partial_v\Phi^{(n-n')}\partial_uD\Phi^{(n')}\big)=0	\,,\label{ggg12}
\end{align}
where, since $\Phi$ is a real field, we must have $\Phi^{(-n)} = (\Phi^{(n)})^*$. The components $\Phi^{(n)}(\tau,u,v)$ can be further expanded in Fourier integral as 
\begin{align}
	\Phi^{(n)}= 
	\int \frac{dk_udk_v}{(2\pi )^2} 
	\xi^{(n)}_{\textbf{k}}(\tau) 
	\exp[i(k_u u+k_v v)]
	\, , 
	\label{ggg13}
\end{align} 
where $\textbf{k}=(k_u,k_v)$.

According to the idea that the system dynamics lives near the axial symmetry, $\Phi^{(0)}$ is expected to be the dominant contribution to the electric field. In the summation terms of Eq.(\ref{ccm10}) and Eq.(\ref{ggg12}), we can henceforth neglect the quadratic terms in $\Phi^{(n\neq0)}$ (we retain only those components in which or $n-n'$ or $n'$ are zero). In this approximation scheme, the Fourier representation of Eq.(\ref{ccm10}) takes the form
\begin{align}
	k^2\partial_{\tau}\xi^{(0)}_{\textbf{k}} 
	- \bar{\rho}_i^2 \int \frac{dq_udq_v}{2\pi}  
	( k_uq_v-k_vq_u) q^2 \big[ 
	\xi^{(0)}_{\textbf{k}-\textbf{q}} 
	\xi^{(0)}_{\textbf{q}}\big] = 0
	\, ,
	\label{ggg14}
\end{align}
while Eq.(\ref{ggg12}) rewrites
\begin{align}
&k^2\partial_{\tau}\xi^{(n)}_{\textbf{k}}+n^2\left( k^2/\bar{\nu}_{ei} + \gamma\right)\xi^{(n)}_{\textbf{k}}-\nonumber\\ 
&- \bar{\rho}_i^2\!\!  \int\!\! \frac{dq_udq_v}{2\pi} \left( 
k_uq_v\!-\!k_vq_u\right) q^2\big[ 
\xi^{(n)}_{\textbf{k} - \textbf{q}} \xi^{(0)}_{\textbf{q}}+ \xi^{(0)}_{\textbf{k} - \textbf{q}} 
\xi^{(n)}_{\textbf{q}} \big] =0\, ,
\label{ggg15}
\end{align}
where, $k^2\equiv k_u^2+k_v^2$ and $q^2\equiv q_u^2+q_v^2$. 
In this scheme, thus linearizing the system around 
the zeroth mode,
we note that Eq.(\ref{ggg14}) becomes the standard 2D
electric turbulent dynamics (corresponding to a 2D Euler equation for an 
incompressible fluid) and its 
properties have been extensively discussed in \citer{montani-fluids2022}.

\subsection{Relevant solution}
It is worth noting that in \citer{montani-fluids2022}, it  was stressed the existence of  a key difference between the fluid and electric dynamics, due to the fact that the former remains valid 
for arbitrarily large value of the wave number modulus 
(i.e. for an arbitrarily small scale), while 
the latter has a natural cut-off scale below the 
wave number associated to the ion Larmor 
radius. We stress that this cut-off value would be, in principle, affected also by magnetic curvature and diamagnetic drifts, \emph{a priori} neglected in the basic dynamical equations of the present model (for a dsicussion of how the diamagnetic effect can induce a relaxed profile of the edge plasma pressure, see \citer{cf22}). The presence of such a cut-off has the following important consequence (easily verifiable 
passing to polar coordinates in the wave number space, see \ref{appA}): 
the 2D electric potential field dynamics 
admits a constant vorticity as steady spectrum, 
i.e. we have the stationary Fourier 
solution
\begin{align}
	\xi^{(0)}_k = \Gamma/k^2
	\, , 
	\label{ggg16}
\end{align}
where $\Gamma$ is a complex constant.
This solution corresponds to 
the statistical equilibrium 
spectrum, derived in \citer{kraichnan67} and \citer{kraichnan75}, 
when the condensation phenomenon is absent, see \citer{seyler75}. Thus, Eq.(\ref{ggg15}) takes, 
around the solution (\ref{ggg16}), the form
\begin{align}
&k^2\partial_{\tau}\xi^{(n)}_{\textbf{k}}+n^2\left( k^2/\bar{\nu}_{ei} + \gamma\right) \xi^{(n)}_{\textbf{k}}-\nonumber\\
&\!\!-\bar{\rho}_i^2\Gamma\!\! \int\!\! \frac{dq_udq_v}{2\pi} \left( 
	k_uq_v\! -\! k_vq_u\right)q^2\Big[ \frac{\xi^{(n)}
	_{\textbf{k} - \textbf{q}}}{q^2} + \frac{\xi^{(n)}_{\textbf{q}}}{(\textbf{k}-\textbf{q})^2} 
\Big] =0\, .
\label{gggx}
\end{align}
This equation admits, as an exact solution, the decaying branch (see \ref{appA}): 
\begin{align}
	\xi^{(n)}_k = \Theta\exp \big[ 
	-n^2\big( (1/\bar{\nu}_{ei}) + \gamma/k^2\big) \tau\big] \, ,
	\label{gggx2}
\end{align}
where $\Theta$ is an assigned complex  constant. The sensitivity of the decaying behavior to
dissipation suggests that the obtained
feature could be affected by additional
parallel contribution and, in particular,
by the magnetic field inhomogeneity,
responsible for a $\textbf{E}\times\textbf{B}$ velocity, which is
no longer divergenceless. However the
interest for this decaying behavior is
nonetheless motivated because it is
induced by the basic ingredients of
non linear drift response. We also note that, substituting the 
expression above for $\xi_k^{(n)}$ into the Fourier transformed summation of Eq.(\ref{ccm10}), it identically vanishes (as it can be easily seen by using polar coordinates in the wave number space and by truncating the domain as in \ref{appA}). In this sense, we have constructed an exact solution of the Fourier transform of the (non linearized) Eq.(\ref{ccm10}).

Concluding, the present study demonstrates the existence 
of a branch of the 3D turbulence 
for which the 2D steady spectrum 
(characterized by the suppression of the condensation 
process) behaves as an attractor. Since we neglected here the linear instability 
trigger, corresponding to the background density 
gradient, this result applies to a self-sustained 
(non-linearly generated) turbulent dynamics; but it resembles the mechanism studied in \citer{biskamp95}, by which the energy of the mode $n\neq0$ is partially transferred back to 2D fluctuations.

\section{Numerical analysis and phenomenological considerations}\label{secnum}
In this Section, we start by showing the general validity of the previous analysis, i.e. that $\Phi^{(0)}$ is the dominant contribution to the electric field, by a direct numerical integration of Eqs.(\ref{ccm10}) and (\ref{ggg12}). In this respect, such a system can be easily rewritten by means of discrete Fourier series expansion in the truncated $k$-space (which would correspond to assign periodic boundary conditions to the problem). In this scheme, $k_u$ and $k_v$ of Eq.\reff{ggg13} become integer numbers, positive and negative, but not both simultaneously zero. For the reality of the field $\xi^{(0)}_{\textbf{k}}$, reversing the sign of $k_u$ or $k_v$ (or both) for this component results into complex conjugation, while we have $\xi^{(-n)}_{\textbf{k}}=(\xi^{(n)}_{\textbf{k}})^*$ for $n\neq0$.

The numerical scheme for integrating the resulting truncated formulation includes a 4th order Runge–Kutta algorithm for the time evolution of the spectral components $\xi^{(n)}_{\textbf{k}}$. For the treatment of the non-linear terms in Eqs.(\ref{ccm10}) and (\ref{ggg12}), we retain the full-$k$ matrix representation of the problem without implementing the often used pseudo-spectral scheme, thus avoiding aliasing issues. We implement physical parameters specified for typical Tokamak machines: $T_{0}=100$eV, $B_0=3$T and $\bar{n}_0=5\times10^{19}$m$^{-3}$ \cite{dtt19}. We thus get $\Omega_i\simeq1.4\times 10^8$s$^{-1}$, while the ion Larmor radius results $\rho_i\simeq0.05$cm. We also consider the poloidal 2D periodicity box length $L$ of the order of the centimeter, while $R\simeq1300$cm. As previously stressed, the physical cut-off at small spatial scales is provided by the ion Larmor radius and, here, we assume such a cut-off at $2\rho_i$ (we note that the parallel component is negligible due to the relation $R\gg L$). We limit the simulations to the components $n=-5,\,...\,,5$ and the system is initialized with a dominant $n=0$ component, having a random vorticity profile ($\propto k^2|\xi_k^{(0)}|$) bounded up to 0.01, which mimics and generalize the solution \reff{ggg16}. The other $|\xi^{(n\neq0)}_{k}|$ components are randomly initialized 2 order of magnitude smaller. Using this setup the initial maximum value for the electrostatic fluctuations is $e\phi\sim 1$eV.
\begin{figure}
\centering
\includegraphics[width=4.3cm]{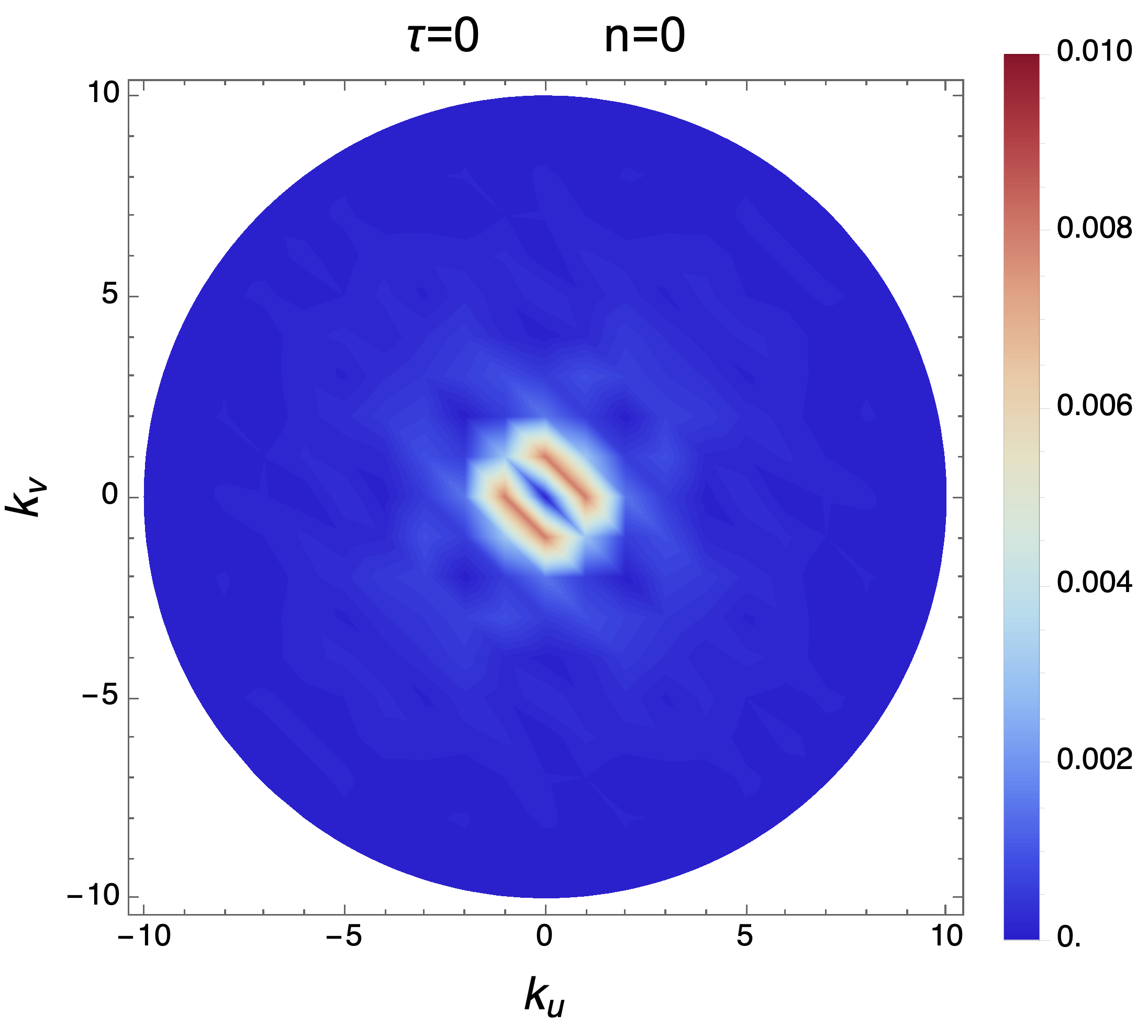}\!\!
\includegraphics[width=4.3cm]{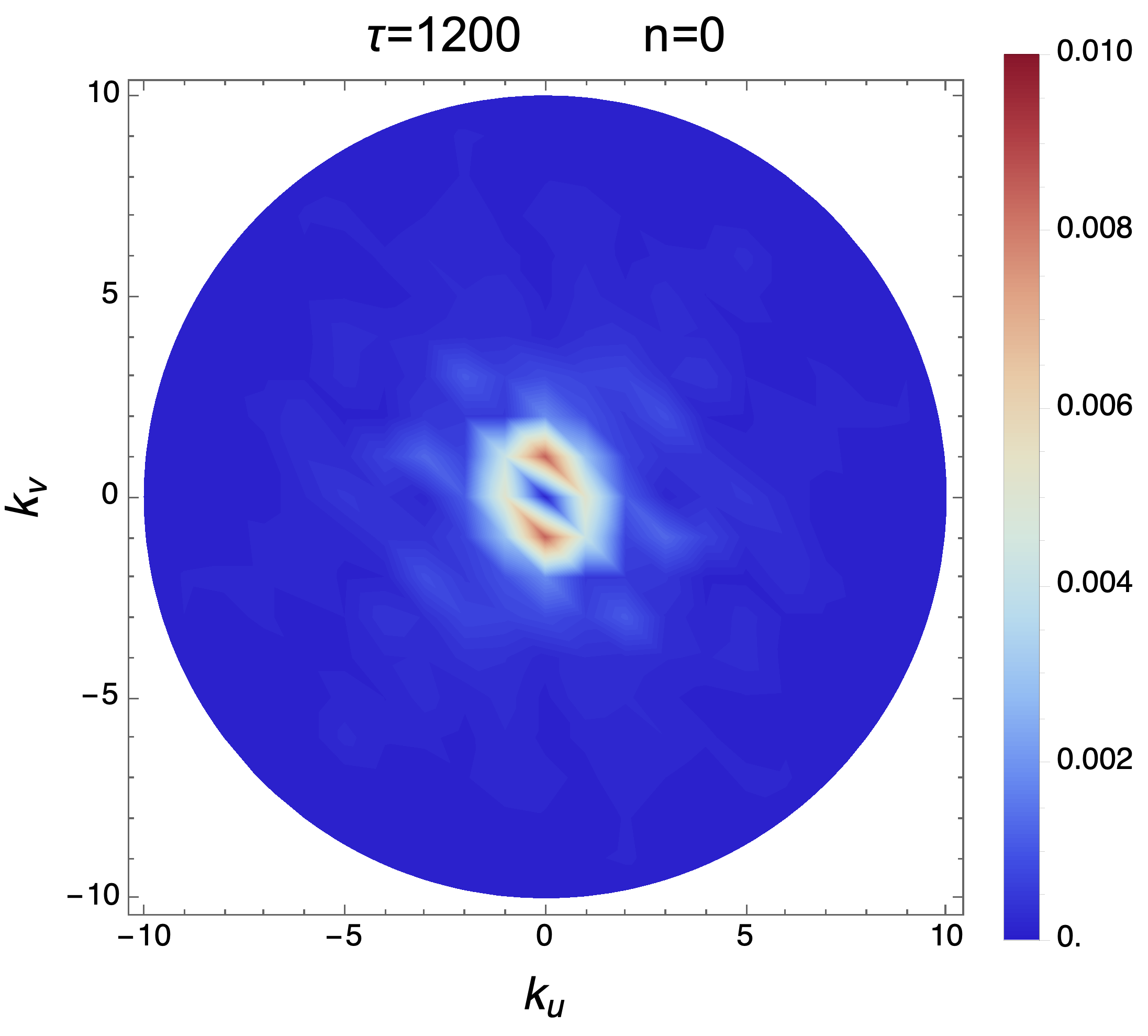}\\
\includegraphics[width=4.3cm]{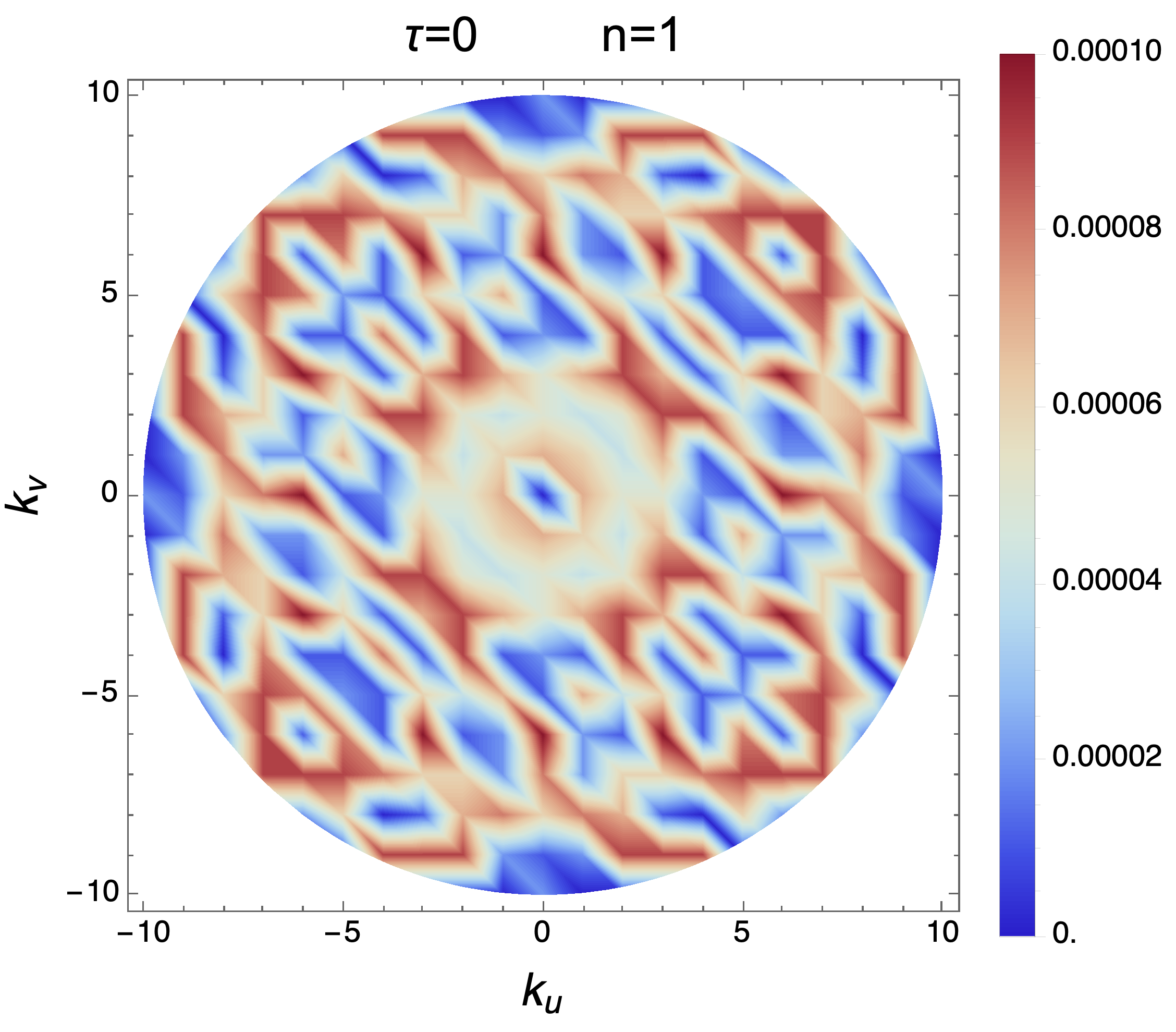}\!\!
\includegraphics[width=4.3cm]{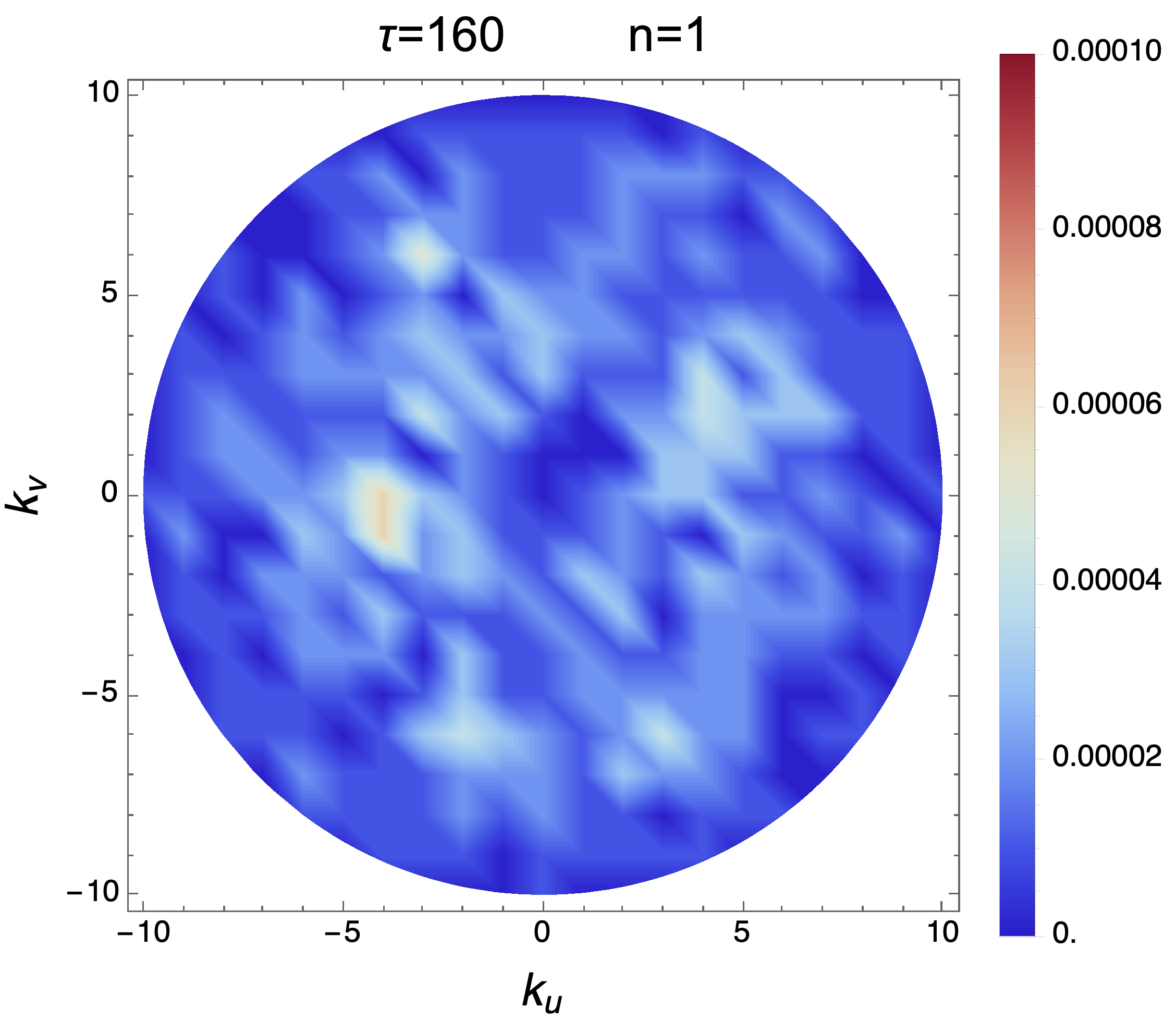}\\
\includegraphics[width=4.3cm]{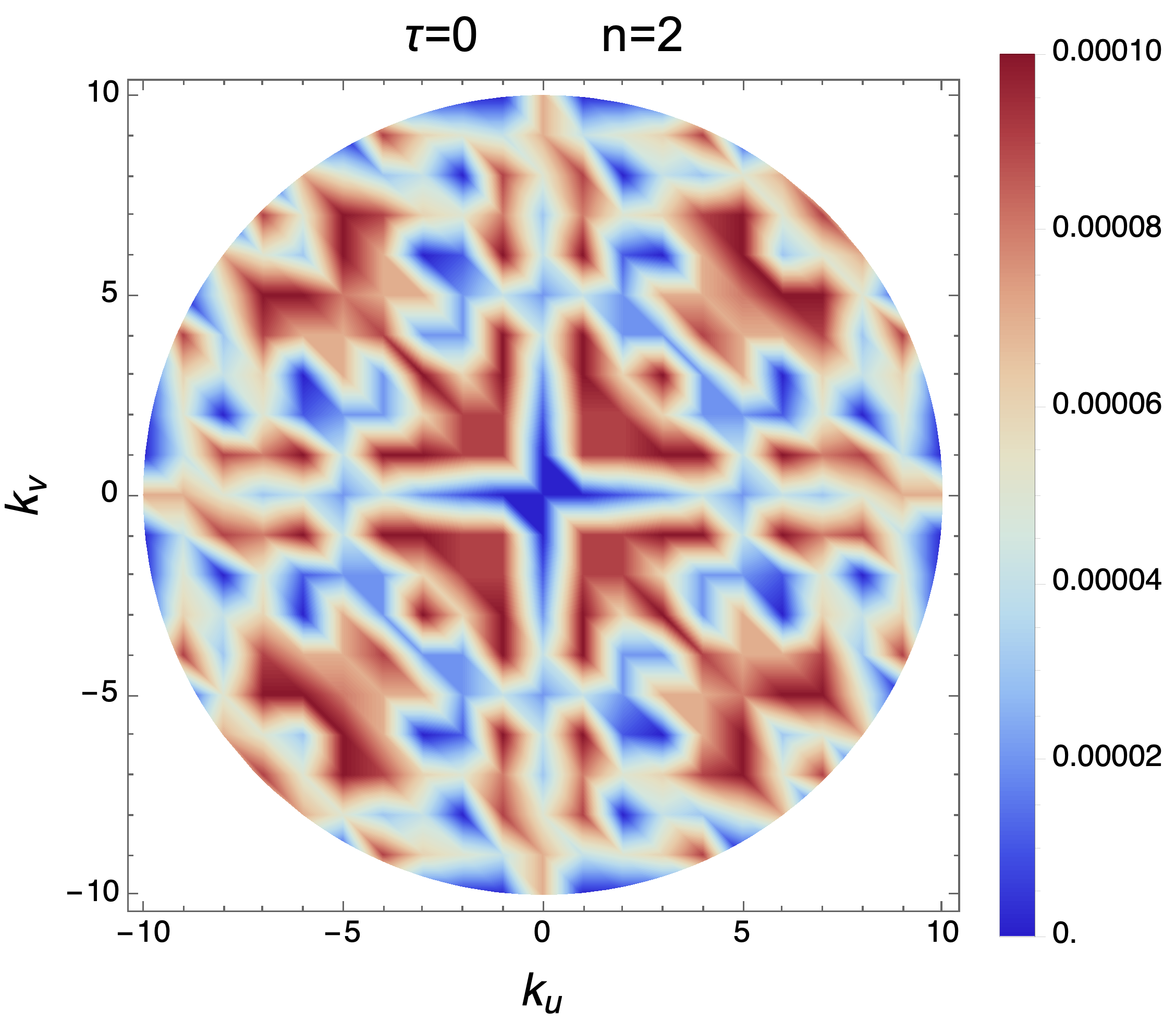}\!\!
\includegraphics[width=4.3cm]{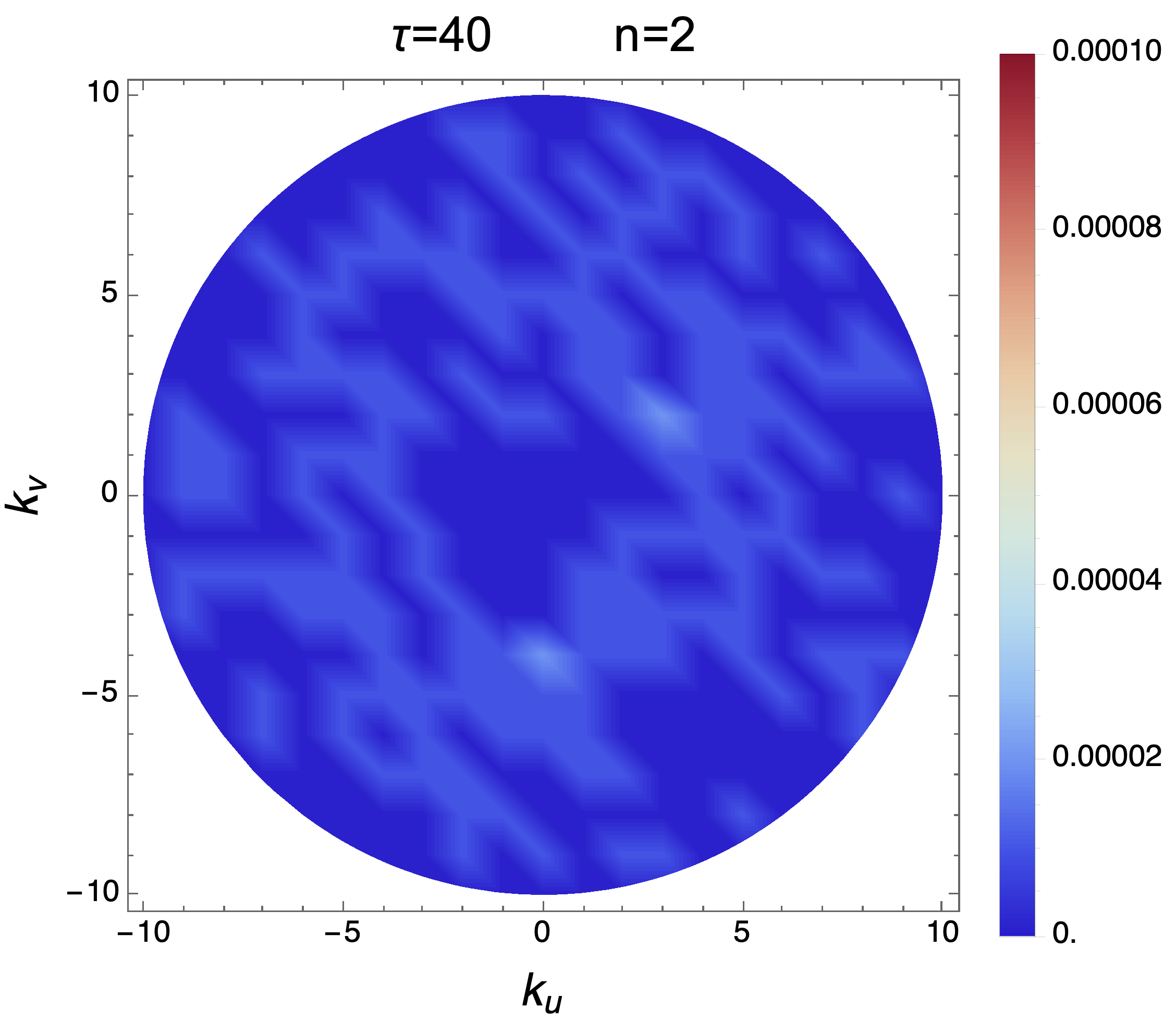}
\caption{Contour plot of $|\xi^{(n=0,1,2)}_{k_u,k_v}|$ (arbitrary units - note the different scales) at several $\tau$, as indicated over the graphs. The data are obtained by numerically integrating Eqs.(\ref{ccm10}) and (\ref{ggg12}) using a discrete Fourier series expansion in the truncated $k$-space. The decaying nature of the $n\neq0$ modes is evident and fast (depending on the toroidal number under consideration).
\label{fig3D}}
\end{figure}

In Fig.\ref{fig3D}, we show the contour plot (in arbitrary units, note the different scales) of $|\xi^{(n)}_{\textbf{k}}|$ at the begin of the simulation and after fixed times, as indicated over the graphs. It clearly emerges how the $n=0$ component, which is initially dominant, is the only surviving one. We show only the components $n=1,\,2$, since the other $n\neq0$ modes behave accordingly. This numerical analysis indicates that the decaying branch (Eq.(\ref{gggx2})) is actually a general solution of the system when considering the $n=0$ modes as initially dominant via their energy contributions. 

Let us now analyze the spectral properties of the only surviving components $\xi^{(0)}_{\textbf{k}}$. Recalling that in a pure 2D scheme the mode energy is isotropic in the $\textbf{k}$ space, we now define the zeroth component spectral energies as 
\begin{align}
W^{(0)}_{k}=k^2|\xi^{(0)}_{k}|^2\,.
\end{align}
\begin{figure}[ht!]
\centering
\includegraphics[width=6cm]{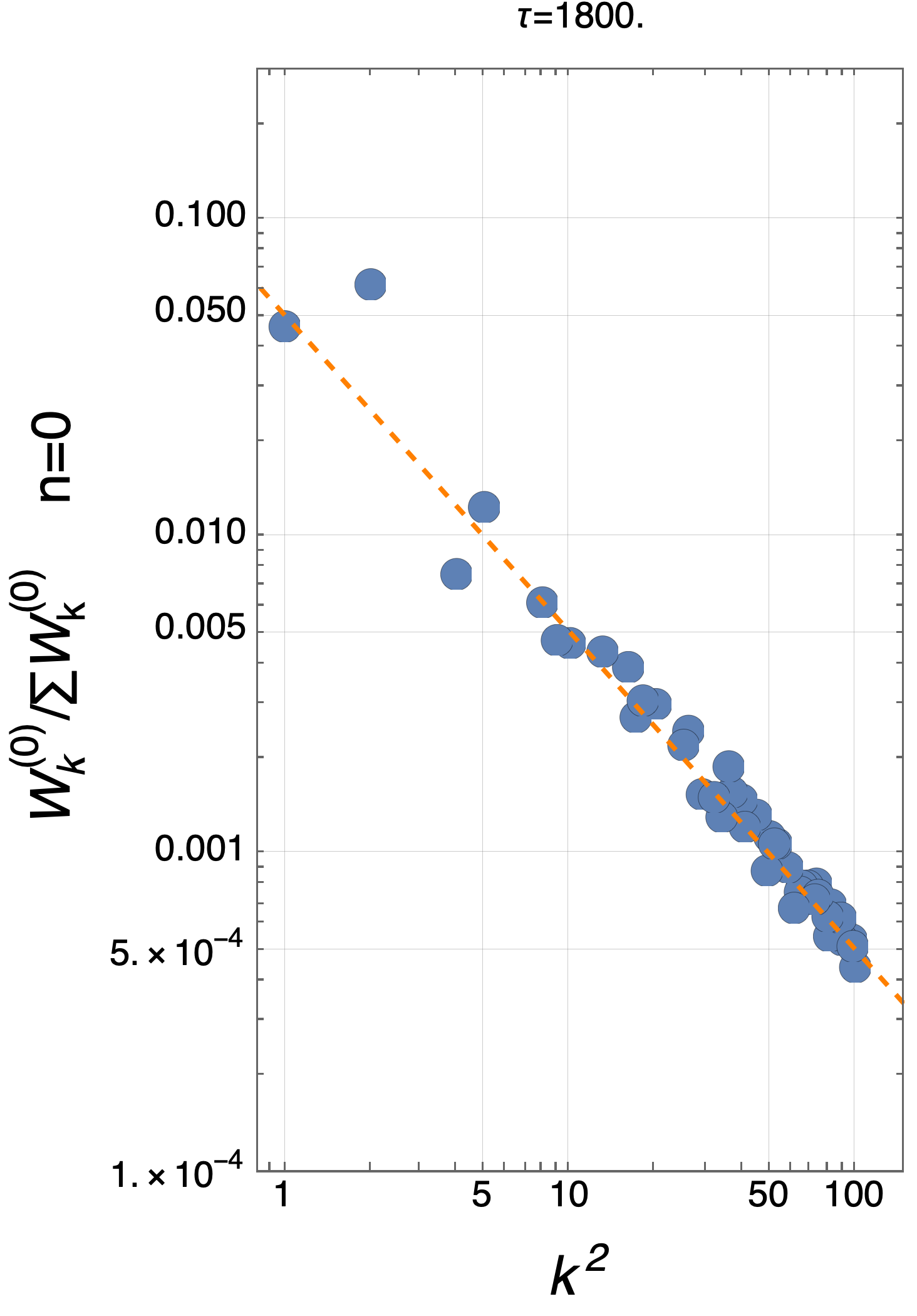}
\caption{Log-log plots of normalized $W^{(0)}_k$ (averaged over 300 time units) as a function of $k^2$. The chose time $\tau=1800$ corresponds to the thermal equilibrium. Orange dashed line indicates the behavior $\sim1/k^2$.
\label{fig2}}
\end{figure}
A single $W_{k}$ is provided by averaging over the pairs $(k_u,k_v)$ yielding the same value of $k$, and we also remark how the evaluated energy spectrum is not strictly instantaneous, in the sense that $W_k$ is time averaged over 300 $\tau$ to avoid fast statistical fluctuations. The energy spectrum analysis is outlined in Fig.\ref{fig2}. We plot the normalized mode energy function and the chosen time corresponds to the thermal equilibrium in the sense that no deviations of the spectrum occurs if we let the system evolve. The constant vorticity steady spectrum solution in Eq.(\ref{ggg16}) implies that 
$W^{(0)}_k\propto k^{-2}$ and it emerges from simulations how the system indeed collapses toward such a profile.

\begin{figure}[ht!]
\centering
\includegraphics[width=7cm]{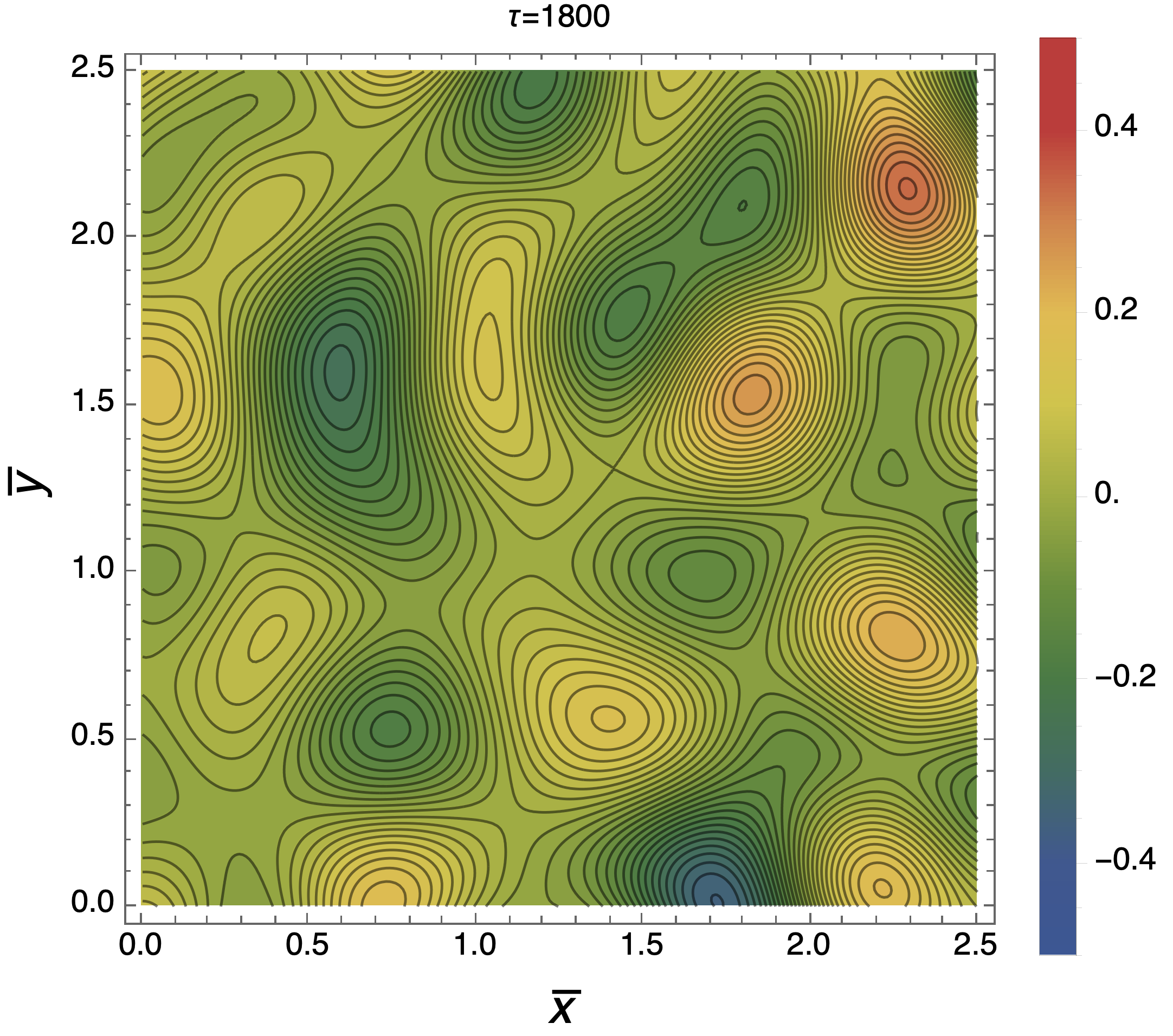}
\caption{Contour plot (in arbitrary units) at $\tau=1800$ of the inverse Fourier transform of $-k^2\xi_{\textbf{k}}^{(0)}$, i.e. $\Delta\Phi^{(0)}(u,v)$.
\label{fig3}}
\end{figure}

Considering the constitutive relation (\ref{ggm3}), which implies $\bar{n}\propto\Delta \Phi$, the behavior of the plasma density is directly related to that one of the 
electric vorticity. In its Fourier 
representation, Eq.(\ref{ggm3}) implies 
that also the turbulent plasma number density 
has a steady constant spectrum $\bar{n}^{(0)}_k$, 
which explicitly reads
\begin{align}
	\bar{n}_k^{(0)} \propto  - k^2 \Phi^{(0)}_k \sim\Gamma	\, . 
	\label{gggf}
\end{align}
This flat spectrum suggests that the plasma 
density is localized in specific regions of 
the physical space, i.e. we deal with small scale blobs and depressions. In Fig.\ref{fig3}, we present a contour plot (taken at the same time of Fig.\ref{fig2}) of the function $\Delta\Phi^{(0)}(u,v)$ constructed by inverse Fourier transforming $-k^2\xi_{\textbf{k}}^{(0)}$. As expected by its constant Fourier spectrum, the density fluctuation (proportional to the profile plotted in the figure) outlines small scale structures. In particular, with respect to the constant background number density $\bar{n}_0$, the electrostatic turbulent behavior induces blobs and depressions at the cut off scale. As discussed in \citer{morel21}, these small scale structures could have important implications in the particle diffusive transport.

As concluding general remark, it is useful to recall that the small wave-number region of the 2D energy spectrum could be characterized, under suitable initial condition, by the condensation phenomenon. Furthermore, in the presence of a small viscosity contribution (always present in a real system), a logarithmic correction must also be considered and the energy is efficiently transferred to large scale structures \cite{kraichnan80,montani-fluids2022}. The formation of large scale vortices should benefit the transport processes, attenuating the diffusivity of particles.

\section{The role of the X-point geometry}\label{secxp}

In order to better characterize the 
model proposed above in view of the 
physics of a Tokamak machine, we introduce, according to \citer{montani-fluids2022}, 
the geometry of a magnetic field X-point. Then, we discuss the resulting impact on 
the dynamics associated to Eq.(\ref{ggm6}).
Due to the stability of the 
$n=0$ mode, theoretically and numerically discussed above, we study the X-point morphology retaining axial symmetry, 
i.e. limiting our attention to the 2D problem only.

We consider a background magnetic configuration of the form
\begin{equation}
	\textbf{B}_0 =B_0\hat{\textbf{b}}= B_p\left( 
	y\hat{\textbf{e}}_x + x\hat{\textbf{e}}_y\right) + B_t\hat{\textbf{e}}_z
	\, , 
	\label{tt1}
\end{equation}
where $B_p$ and $B_t$ are two constants with $B_p\ll B_t$. If we introduce the small parameter $ \varepsilon=B_pL/B_t \ll 1$, the magnetic versor $\hat{\textbf{b}}$ reads
\begin{equation}
\hat{\textbf{b}} = \frac{1}{\sqrt{1 + \varepsilon^2\left( u^2 + v^2\right)}}\left( \varepsilon v\, ,\, \varepsilon u\, ,\, 1\right) \simeq \left( \varepsilon v\, ,\, \varepsilon u\, ,\, 1\right)\, ,
	\label{tt2}
\end{equation}
where we approximated the expression 
at first order in $\varepsilon$.

In what follows, we always retain 
terms at the lowest order in the parameter $\varepsilon$, to 
account for the X-point geometry 
as a perturbation to the axial 
magnetic field in the $z$-direction, i.e. all the quantities of the background and the perturbations are now not dependent on $w$. It is however important to stress 
that the suppression of the $w$-derivatives does not imply, due to the X-point geometry, to deal with zero parallel derivatives. In fact, we have 
the following relation: 
\begin{equation}
	\hat{\textbf{b}}\cdot \nabla \equiv \partial_{\parallel} = \varepsilon 
	\left( v\partial_u + u\partial_v\right)	\, .
	\label{tt3}
\end{equation}
Hence, we get the dimensionless parallel and orthogonal Laplacian in the form
\begin{equation}
	\Delta_{\parallel} = \partial_{\parallel}^2 \,,\qquad 
	\Delta_{\perp} = 
	(\nabla - \hat{\textbf{b}}\partial_{\parallel})^2 =
	D - \partial_{\parallel}^2 + \partial\equiv\mathcal{D}\, , 
	\label{tt4}
\end{equation}
where $\partial\equiv \varepsilon^2\left(
u\partial_u + v\partial_v\right)$.

Now, observing that in the present scheme the $w$-derivative is replaced by $\partial_{\parallel}$, and $D$ 
by $\mathcal{D}$ (which explicitly reads $\mathcal{D}= D - \varepsilon^2 (v^2\partial_u^2 + u^2\partial _v^2 + 2uv\partial_u\partial _v)$), we can easily infer that Eq.(\ref{ggm6}) rewrites as
\begin{align}
	\partial_{\tau}\mathcal{D}\Phi 
	+ \tilde{\rho}_i^2( \partial_u\Phi \partial_v\mathcal{D}\Phi&- \partial_v\Phi \partial_u\mathcal{D}\Phi) =\nonumber\\
	&=\partial^2_{\parallel}\left((1/\bar{\nu}_{ei}) D\Phi - \gamma\Phi\right)
	\, ,
	\label{sfs1}
\end{align}
where $\tilde{\rho}_i^2\equiv \bar{\rho}_i^2(1 - \varepsilon^2( 
u^2+v^2))$ (this factor comes from the 
first order expansion in $\varepsilon^2$ of the $\textbf{E}\times\textbf{B}$ velocity). Above, we intend to retain the zeroth and 
first order terms in $\varepsilon^2$ only.

The feature on which we focus our 
attention is that the model here proposed admits, in the presence of the 
X-point morphology, the following unstable exact solution:
\begin{equation}
	\Phi (\tau ,u,v) = e^{2\gamma \tau}uv 
	\, .
	\label{sfs2}
\end{equation}
The physical interpretation of this 
electric profile is that of a ``local'' 
(not constrained by the boundary conditions) burst near the X-point. 
This issue suggests that the stability 
of the 2D turbulence 
profile is deeply altered in a 
region sufficiently close to the 
null configuration of a Tokamak device, see \citer{tcv21} for experimental evidences in this direction.

\section{Concluding remarks}

We constructed a reduced turbulence model, having the 
basic features of a Hasegawa-Wakatani theory, 
in which the linear instability trigger (due 
to the number density gradient) and the 
ion viscosity have been removed.
This reduction of the dynamics allows 
to establish a constitutive relation 
between the electric field Laplacian and 
the number density fluctuation. 
It is worth stressing that, although such 
a relation resembles a Gauss law, 
no charge density is here present since 
the quasi-neutrality of the plasma has been assumed at the ground of the present model.

The proposed scheme has a 3D structure
summarized by a single equation for the 
electric field potential, in which 
perpendicular and parallel derivatives 
to the background magnetic field appear. We investigated the profile of 
the turbulence spectrum remaining 
close to the axial symmetry of a 
Tokamak, i.e. linearizing the dynamics 
with respect to the $n\neq 0$ Fourier harmonics. 
The behavior of the $n=0$ mode was instead 
chosen as the corresponding 2D steady spectrum.

The most important result of our analysis consisted in showing the existence 
of a decaying (stable) branch of the 
3D dynamics, for which the 2D spectrum 
behaves as an attractor. This issue was also confirmed by a 3D numerical analysis in the correspondence to initial conditions in which the $n=0$ mode dominates. 
Since our reduced model does not 
contain the background number density 
gradient, which corresponds to the 
linear instability growth rate, 
our result acquires a specific significance 
in the spirit of the results 
presented in \citers{scott90,scott02}. 
In fact, there it is stated that the non-linear 
drift response is a purely non-linear 
self-sustained phenomenon, no requiring 
the linear triggering. 
The idea suggested by our study is that 
the basic ingredient of this non-linear 
drift response be the electrostatic 2D turbulence, investigated in \citer{montani-fluids2022}, which has a direct 
mapping with the turbulence proper of 
an incompressible fluid \cite{kraichnan80}. Moreover, we discussed how the introduction of the X-point geometry affects the considered dynamics. In the resulting scheme, we outlined the existence of a non-linear burst-like solution which could shed light on the real nature of the X-point turbulence.

The analogy stressed above between the Euler equation of an
incompressible fluid 2D dynamics and the electrostatic turbulence can be easily understood, from a physical point of view, by observing that, for a (nearly) uniform magnetic field, the $\textbf{E}\times\textbf{B}$ velocity of the ion fluid is divergenceless. However, as discussed in Sec.\ref{secxp}, small inhomogeneities near the X point of a Tokamak configuration do not significantly affect the ion incompressibility.

Finally, the considered reduced model 
allows to directly infer the steady 
configuration of the number density
fluctuations (as viewed in the physical 
space) and we outlined a distribution 
of small scale blobs and depressions, 
analogously to more general settings \cite{numata07}. 
Since this last result came out 
from having neglected the background density 
gradient, it constitutes a further 
indication that the electrostatic turbulence is a non-linear 
self-sustained phenomenon.

\appendix
\section{}\label{appA}
In order to derive the solutions 
(\ref{ggg16}) and (\ref{gggx2}), the 
following relations are concerned. Let us consider the integral
\begin{align}
	\int_\Sigma dq_udq_v
	\frac{k_uq_v-k_vq_u}{(\textbf{k}-\textbf{q})^2}
	\, , 
	\label{a1}
\end{align}
where, without any truncation in the 
wave number space, the domain $\Sigma$ is 
the whole 2D plane. If we now consider the coordinate 
substitution $\textbf{q}^{\prime} \equiv \textbf{k} - \textbf{q}$, the integral above rewrites
\begin{align}
	\int_{\Sigma^{\prime}} dq^{\prime}_udq^{\prime}_v\frac{k_uq^{\prime}_v-k_vq^{\prime}_u}{(\textbf{q}^{\prime})^2} 
	\, , 
	\label{a2}
\end{align}
where clearly $\Sigma^{\prime}\equiv \Sigma$, since 
$\textbf{q}^{\prime}$ still spans the 
whole plane. In order to truncate the $k$-space, taking into account the isotropy of the plasma (and hence of the spectrum), we must restrict $\Sigma$ to the domain $\tilde{\Sigma}$, which is delimited by the 
circumference of radius $k_{max}$, 
(which is the cut-off value of the 
wave number modulus). 
It is therefore natural to restrict 
the domain $\Sigma^{\prime}$ to the same 
region $\tilde{\Sigma}$. Let us now introduce polar coordinates $\textbf{q}^{\prime} = (q^{\prime}\cos \varphi,\;q^{\prime}\sin\varphi)$ 
(analogously $\textbf{k} = (k\cos\theta,\;k\sin\theta)$). The integrals (\ref{a1}) and (\ref{a2}) become
\begin{align}
&\int_{\tilde{\mathcal{D}}}dq^{\prime}d\varphi k(\cos\theta\sin\varphi-
\sin\theta\cos\varphi)=\nonumber\\
&=kk_{max}\int_0^{2\pi}d\varphi(\cos\theta\sin\varphi-\sin\theta\cos\varphi)
\equiv 0\, , \label{a3}
\end{align}
where we note that, in the integral above, $\theta$ is a constant angle.

\vspace{1cm}
\textbf{Acknowledgments} We would like to thank Dr. Fabio Moretti for discussions on this topic and contributions to Sec.\ref{secxp}.


\end{document}